\documentclass[12pt,preprint]{aastex}
\usepackage{url} 
\usepackage{amsmath}
\usepackage{epsfig}
\usepackage{graphicx}
\usepackage{color}
\newcommand\lsim{\mathrel{\rlap{\lower4pt\hbox{\hskip1pt$\sim$}}
        \raise1pt\hbox{$<$}}}
\newcommand\gsim{\mathrel{\rlap{\lower4pt\hbox{\hskip1pt$\sim$}}
        \raise1pt\hbox{$>$}}}

\newcommand{\dd}{\partial}

\defcitealias{MP05}{MP05}
\defcitealias{LP74}{LP74}
    \def\dd{\partial}

    \def\beq{\begin{equation} }
    \def\eeq{\end{equation} }
    \def\spose#1{\hbox to 0pt{#1\hss}}
    \def\ltsim{\mathrel{\spose{\lower.5ex\hbox{$\mathchar"218$}}
     \raise.4ex\hbox{$\mathchar"13C$}}}

\def\spose#1{\hbox to 0pt{#1\hss}}
\def\lta{\mathrel{\spose{\lower 3pt\hbox{$\mathchar"218$}}
        \raise 2.0pt\hbox{$\mathchar"13C$}}}
\def\gta{\mathrel{\spose{\lower 3pt\hbox{$\mathchar"218$}}
        \raise 2.0pt\hbox{$\mathchar"13E$}}}

\begin{document}
\title
{Exact Time-dependent Solutions for the Thin Accretion Disc Equation: Boundary Conditions at Finite Radius}
\author{Takamitsu Tanaka}
\affil{Department of Astronomy, Columbia University, 550 West 120th Street, New York, NY 10027}
\shorttitle{Green's Functions for Thin Discs with Finite Boundaries}
\shortauthors{T. Tanaka}
\begin{abstract}
We discuss Green's-function solutions of
the equation for a geometrically thin, axisymmetric Keplerian accretion disc
with a viscosity prescription $\nu\propto R^{n}$.
The mathematical problem was solved by \cite{LP74}
for the special cases with boundary conditions
of zero viscous torque and zero mass flow at the disc center.
While it has been widely established that the observational appearance
of astrophysical discs depend on the physical size of the central object(s),
exact time-dependent solutions with boundary conditions imposed
at finite radius have not been published for a general value
of the power-law index $n$.  We derive exact Green's-function solutions that
satisfy either a zero-torque or a zero-flux condition at a nonzero inner boundary $R_{\rm in}>0$,
for an arbitrary initial surface density profile.
Whereas the viscously dissipated power diverges at the
disc center for the previously known solutions with $R_{\rm in}=0$,
the new solutions with $R_{\rm in}>0$ have finite expressions
for the disc luminosity that agree, in the limit $t\rightarrow \infty$,
with standard expressions for steady-state disc luminosities.
The new solutions are applicable to the evolution of the innermost regions of thin accretion discs.
\end{abstract}
\keywords{
accretion, accretion discs
}

\section{Introduction}

Since its emergence in the 1970's \citep{SS73, NT73, LP74},
the theory of astrophysical accretion discs has been applied
to explain the emission properties of active galactic nuclei,
X-ray binaries, cataclysmic binaries, supernovae, gamma-ray bursts,
and the electromagnetic signatures of mergers of supermassive black holes;
to study planetary and star formation;
and to model the evolution of binary and planetary systems.
Because accretion discs onto compact objects can dissipate much larger fractions
of baryonic rest-mass energies than nuclear reactions,
they are often associated with some of the most energetic
astrophysical processes in the universe.

If the local gravitational potential is dominated by a central
compact object or a compact binary,
and if the timescale for the viscous dissipation of energy is
longer than the orbital timescale,
then the accretion flow near the center of the potential
is expected to be nearly axisymmetric.
If the gas is able to cool efficiently, then the flow will also be geometrically thin,
and one only needs the radial coordinate to describe the mass distribution in the disc
(any relevant vertical structure can be integrated or averaged over the disc height).
The partial differential equation \citep[][henceforth LP74]{LP74}
\beq
\frac{\dd}{\dd t}\Sigma (R,t)=\frac{1}{R}\frac{\dd}{\dd R}\left[ R^{1/2}\frac{\dd}{\dd R}\left(3\nu\Sigma R^{1/2}\right)\right],
\label{eq:diff}
\eeq
is obtained by combining the equations of
mass conservation and angular momentum,
and describes the surface density evolution of a thin Keplerian accretion disc
due to kinematic viscosity $\nu$.

In general, the viscosity $\nu$ depends on the surface
density $\Sigma$ and equation \eqref{eq:diff} is nonlinear.
If, however, $\nu$ is only a function of radius, then the equation
is linear and much more amenable to analytic methods.
In particular, a solution that makes use of a Green's function $G$,
\beq
\Sigma (R,t)=\int_{R_{\rm in}}^{\infty} G(R,R^{\prime}, t)~\Sigma (R^{\prime}, t=0)~dR^{\prime},
\label{eq:Green}
\eeq
gives the solution $\Sigma$ for any $t>0$ given an arbitrary
profile $\Sigma (R,t=0)$
and an inner boundary condition imposed at $R_{\rm in}$.
A distinct advantage of the formalism is that it gives the solution $\Sigma(R,t)$ through a single
ordinary integral, whereas a finite-difference algorithm would require
the computation of the profile at intermediate times.  Another advantage
is that the initial density profile need not be differentiable.

For a power-law viscosity $\nu\propto R^{n}$,
\cite{Lust52} and \citetalias{LP74} derived analytic Green's functions
that satisfy a boundary condition of either zero-torque or zero-mass-flux
at the coordinate origin, i.e., for the case $R_{\rm in}=0$.
In reality, however, the objects at the center of astrophysical accretion discs
have a finite size to which the observational appearance
of the disc is sensitive: e.g., the luminosity, spectral hardness, and variability timescales
of black hole discs depend strongly on the radius of the innermost stable orbit,
and those of circumbinary discs depend on where the inner
disc is truncated by the central tidal torques.
Green's functions with $R_{\rm in}=0$ do not capture
the time-dependent behavior of accretion discs
close to the central object.  In fact, in solutions obtained with
such Green's functions, the integral for the
total power viscously dissipated in the center of the disc diverges.

Despite the astrophysical relevance of Green's functions to the thin accretion disc equation
with boundary conditions imposed at a finite radius,
such solutions have not been published.
\cite{Pringle91} derived the Green's function with a zero-flux
boundary condition at a nonzero radius in the special
case $n=1$, and noted the ``extreme algebraic complexity''
involved in calculating a more general solution with $R_{\rm in}>0$.
Time-dependent models of accretion flows have continued to 
employ solutions that correspond to the central objects
having zero physical size \citep[e.g., ][]{Metzger+08, TM10}.
In order to calculate a convergent disc luminosity and spectrum,
such models typically approximate analytically the effects of an inner boundary
condition, e.g., by truncating the disc profile at an artificially imposed radius.

In this paper we derive exact Green's functions for equation \eqref{eq:diff} for
boundary conditions imposed at a finite radius, for
any power-law viscosity $\nu\propto R^{n}$ with $n<2$.
We show that mathematical difficulties can be minimized
with the aid of the appropriate integral transform techniques,
namely the Weber transform \citep{Titchm23} and
the recently proved generalized Weber transform \citep{ZT07}.
We present two specific solutions of astrophysical interest:
 the solution with zero torque at a radius $R_{\rm in}>0$,
which is of interest for accretion discs around black holes
and slowly rotating stars;
and the solution with zero mass flow at $R_{\rm in}>0$, which
is applicable to accretion flows that accumulate mass at the disc center
due to the injection of angular momentum from the tidal torques of a binary
or perhaps the strong magnetic field of the central object.

This paper is organized as follows.  In \S 2, we review the 
Green's function solutions, derived by \cite{Lust52} and \citetalias{LP74}, for the thin-disc equation
with boundary conditions imposed at the origin.
In \S 3, we derive the new Green's function solutions,
which impose boundary conditions at a finite inner boundary radius.
We  offer our conclusions in \S 4.

\section{Green's-Function Solutions with Boundary Conditions at $R=0$}
In the special case where the viscosity is a radial power law, $\nu\propto R^{n}$,
and assuming a separable ansatz of the form $\Sigma(R,t)=R^{p}\sigma(R)\exp(-\Lambda t)$,
where $p$ and $\Lambda$ are real numbers and $\sigma$ is an arbitrary function of $R$,
equation \eqref{eq:diff} can be rewritten as the Bessel differential equation:
\beq
R^{2}\frac{\dd^{2}\sigma}{\dd R^{2}}+\left(2p+2n+\frac{3}{2}\right)R\frac{\dd \sigma}{\dd R}
+\left[\left(p+n\right)\left(\frac{\Lambda}{3s}R^{2-n}+p+n+\frac{1}{2}\right)\right]\sigma=0.
\label{eq:Bessel}
\eeq
Above, $s=\nu R^{-n}$ is a constant.
With the choices $p=n-1/4$ and $\Lambda=3sk^{2}$, equation \eqref{eq:Bessel}
has the general solution
\beq
\sigma_{k}(R)=
R^{-2n}\left[ A(k) J_{\ell}(k y) + B(k)Y_{\ell}(k y)\right].
\eeq
Above, $k$ is an arbitrary mode of the solution;
$A(k)$ and $B(k)$ are the mode weights;
$\ell = (4-2n)^{-1}>0$; $y(R)\equiv R^{(1-n/2)}/(1-n/2)$;
and 
$J_{\ell}$ and $Y_{\ell}$ are the Bessel functions of the first and second kinds, respectively, and of order $\ell$.
If $\ell$ is not an integer, then $Y_{\ell}$ above may be replaced without loss of generality by $J_{-\ell}$.
Integrating the fundamental solution across
all possible $k$-modes gives the solution:
\beq
\Sigma (R,t) = \int_{0}^{\infty}R^{-n-1/4}\left[ A(k) J_{\ell}(k y)
+ B(k)Y_{\ell}(k y)\right]~
\exp(-3s k^{2}t)~dk.
\label{eq:gensol}
\eeq

The mode-weighting functions $A(k)$ and $B(k)$ are determined
by the boundary conditions and the initial surface density profile $\Sigma (R,t=0)$.
Our goal is to rewrite equation \eqref{eq:gensol} in
the Green's function form (equation \ref{eq:Green}) and to
write down an explicit symbolic expression for the Green's function $G(R,R^{\prime},t)$.
Throughout this paper, we will employ the following strategy:
\begin{enumerate}
\item Using the boundary condition,
find an analytic relationship between the mode weights $A(k)$ and $B(k)$.
\item Identify the appropriate integral transform to express the mode weights in terms of the initial profile $\Sigma (R,t=0)$.
\item Insert the time-dependence $\exp(-3sk^{2}t)$ and integrate over all modes to find the Green's function.
\item Derive analytic expressions for the asymptotic disc behavior at late times and small radii.
\end{enumerate}
Before deriving the solutions with boundary conditions at
finite radius, we begin by reviewing the Green's functions
of \citetalias{LP74} with boundary conditions at the coordinate origin.

\subsection{Zero torque at $R_{\rm in}=0$}
\label{subsec:zerotorqueR0}
An inner boundary condition with zero central torque is of
astrophysical interest as it can be used to describe accretion
onto a black hole or a slowly rotating star, at radii much larger
than the radius of innermost circular orbit or the stellar surface, respectively.
The radial torque density $g$ in the disc due to viscous shear is
\beq
g(R,t)=\nu\Sigma R^{2}\frac{\dd \Omega_{\rm K}}{\dd R}\propto \nu\Sigma R^{1/2},
\eeq
where $\Omega_{\rm K}$ is the Keplerian angular velocity of the orbit.

Because the functions $J_{\ell}$ and $Y_{\ell}$ have the asymptotic behaviors
$J_{\ell}(ky)\propto y^{\ell}\propto R^{1/4}$ and
$Y_{\ell}(ky)\propto y^{-\ell}\propto R^{-1/4}$
near the origin, at small radii the mode weight $A(k)$
will contribute to the behavior $g\propto R^{1/2}$
while $B(k)$ will contribute to $g={\rm constant}$.
Therefore, for the solution to have zero viscous torque at $R=0$
the function $B(k)$ must be identically zero.

We may relate the surface density distribution at $t=0$ and
the weight $A(k)$ via the integral equation
\beq
\Sigma (R,t=0)=R^{-n-1/4}\int_{0}^{\infty}A(k)J_{\ell}(ky)~dk,
\eeq
which may be solved with the use of the Hankel integral transform \citep[e.g., ][]{Ogilvie05}.

A Hankel transform pair of order $\ell$ satisfies 
\begin{align}
\phi_{\ell}(x)&=\int_{0}^{\infty}\Phi_{\ell}(k)~J_{\ell}(kx)~k~dk,\\
\Phi_{\ell}(k)&=\int_{0}^{\infty}\phi_{\ell}(x)~J_{\ell}(kx)~x~dx.
\end{align}
For the problem at hand, the suitable transform pair is
\begin{align}
R^{n+1/4}\Sigma(R,t=0)&=\int_{0}^{\infty}\left[A(k)k^{-1}\right]J_{\ell}(ky)~k~dk,\\
A(k)k^{-1}&=\int_{0}^{\infty}\left[R^{n+1/4}\Sigma(R,t=0)\right]J_{\ell}(ky)~y~dy.
\end{align}
Combining them gives us $A(k)$:
\beq
A(k)=\left(1-\frac{n}{2}\right)^{-1}\int_{0}^{\infty}
\Sigma(y^{\prime},0)~J_{\ell}(ky^{\prime})\;k\; R^{\prime 5/4} ~dR^{\prime}
\label{eq:sig2}
\eeq

Inserting equation \eqref{eq:sig2} and $B(k)=0$ into equation \eqref{eq:gensol}, we obtain
\beq
\Sigma(R,t)=\left(1-\frac{n}{2}\right)^{-1}R^{-n-1/4}\int_{0}^{\infty}R^{\prime 5/4}\int_{0}^{\infty}
\Sigma(R^{\prime},t=0)~J_{\ell}(ky^{\prime})~J_{\ell}(ky) ~\exp\left(-3sk^{2}t\right)~ k\;dk\;dR^{\prime}.
\eeq
To pose the solution in terms of a Green's function $G(R,R^{\prime},t)$ (equation \ref{eq:Green}),
we write
\begin{align}
G(R,R^{\prime},t)&=\left(1-\frac{n}{2}\right)^{-1}R^{-n-1/4}R^{\prime 5/4}\int_{0}^{\infty}
J_{\ell}(ky^{\prime})~J_{\ell}(ky) ~\exp\left(-3sk^{2}t\right)~ k\;dk\nonumber\\
&=(2-n)\frac{R^{-9/4}R^{\prime 5/4}}{\tau(R)}
I_{\ell}\left[\frac{2\left(R^{\prime}/R\right)^{1-n/2}}{\tau(R)}\right]
\exp\left[-\frac{1+\left(R^{\prime}/R\right)^{2-n}}{\tau(R)}\right].
\label{eq:G1}
\end{align}
Above, $I_{\ell}$ is the modified Bessel function of the first kind,
and we have substituted
$\tau(R)\equiv 12(1-n/2)^{2}R^{n-2}st = 8(1-n/2)^{2}[t/t_{\nu}(R)]$,
where $t_{\nu}(R)=(2/3)R^{2}/\nu(R)$ is the local viscous timescale at $R$.

Although the Green's function allows for the calculation of $\Sigma(R,t)$
for arbitrary initial surface density profiles,
it is instructive to study the case where the initial surface density is
a Dirac $\delta$ function,
\beq
\Sigma (R,t=0)=\Sigma_{0}~\delta(R-R_{0})~R_{0},
\label{eq:delta}
\eeq
for which the solution is (by definition)
the Green's function itself.
The integral over radius
in equation \eqref{eq:Green} becomes trivial and many behaviors
of the solution may be expressed analytically.
Because any initial surface density profile can be described as a superposition
of $\delta$-functions, studying this special case will help illuminate
the general behavior of all solutions.

We may evaluate the asymptotic behavior at late times
and small radii
by noting that for small argument $z\ltsim 0.2\sqrt{1+\ell}$,
$I_{\ell}(z)\approx (z/2)^{\ell}/\Gamma(\ell+1)$.
We find
\beq
\Sigma\left(R, t\ga t_{\nu}(R)\right)\approx
\frac{2-n}{ \Gamma(\ell +1)}\Sigma_{0}
\left(\frac{R}{R_{0}}\right)^{-n}\left[8\left(1-\frac{n}{2}\right)^{2}\frac{t}{t_{\nu,0}}\right]^{-1-\ell},
\label{eq:asym1}
\eeq
where $t_{\nu,0}\equiv t_{\nu}(R_{0})$.

Thus, for these solutions  the inward radial mass flow,
\beq
\dot{M}(R)=-2\pi R\Sigma v_{R}=6\pi R^{1/2}\frac{\dd}{\dd R}\left(\nu\Sigma R^{1/2}\right),
\eeq
becomes radially constant near the origin and at late times:
\beq
\dot{M}\left(t\ga t_{\nu}(R)\right)
\approx \frac{2-n}{ \Gamma(\ell +1)}\dot{M}_{0}
\left[8\left(1-\frac{n}{2}\right)^{2}\frac{t}{t_{\nu,0}}\right]^{-1-\ell}.
\label{eq:Mdot}
\eeq
Above, we have defined $\dot{M}_{0}\equiv 3\pi\nu (R_{0})\Sigma_{0}$.

The power per unit area that is locally viscously dissipated from
each face of the disc is $F=(9/8)\nu\Sigma \Omega^{2}$.
The total power dissipated near the center of the disc diverges:
\beq
L\left(R, t\ga t_{\nu}(R)\right)=\int_{0}^{R}\frac{9\pi}{2}\nu(R^{\prime})\Sigma(R^{\prime},t)\Omega^{2}(R^{\prime}) R^{\prime}~dR^{\prime}
\propto\int_{0}^{R}R^{\prime-2}~dR^{\prime}.
\eeq
Astrophysical accretion flows do not extend
to zero radius, and thus in practice one may truncate the disc
at some plausible boundary radius, for example the radius
of innermost stable circular orbit for a disc around a black hole,
by approximating the effects of a finite boundary radius \citepalias{LP74}.

Although we have used a $\delta$ function for demonstrative purposes,
the quantities $R_{0}$ and $\Sigma_{0}$ that set the
physical scale and normalization of the initial surface density profile,
respectively, are arbitrary.
The asymptotic behaviors noted above hold for
any initial surface density profile: at late times,
the surface density profile
approaches $\Sigma\propto R^{-n}$,
$\dot{M}$ becomes radially constant, 
and the disc luminosity $L$ formally diverges at the center.

At early times and large radii, such that $t\ll \sqrt{t_{\nu}(R) ~t_{\nu}(R^{\prime})}$,
we may use the fact that $I_{\ell}(z\gg1) \approx \exp(x)/\sqrt{2\pi z}$ to find
\beq
G\left(t\ll \sqrt{t_{\nu}(R) ~t_{\nu}(R^{\prime})}\right)\approx
\frac{1}{\sqrt{\pi\tau(R)}}\exp\left\{-\frac{\left[1-\left(R^{\prime}/R\right) ^{1-n/2}\right]^{2}}{\tau(R)}\right\}
\left(\frac{R^{\prime}}{R}\right)^{(3/4)(1+n)}
\frac{d}{dR^{\prime}}\left[\left(\frac{R^{\prime}}{R}\right)^{\prime 1-n/2}\right].
\label{eq:Gearly}
\eeq

\begin{figure}
\centering
\epsfig{file=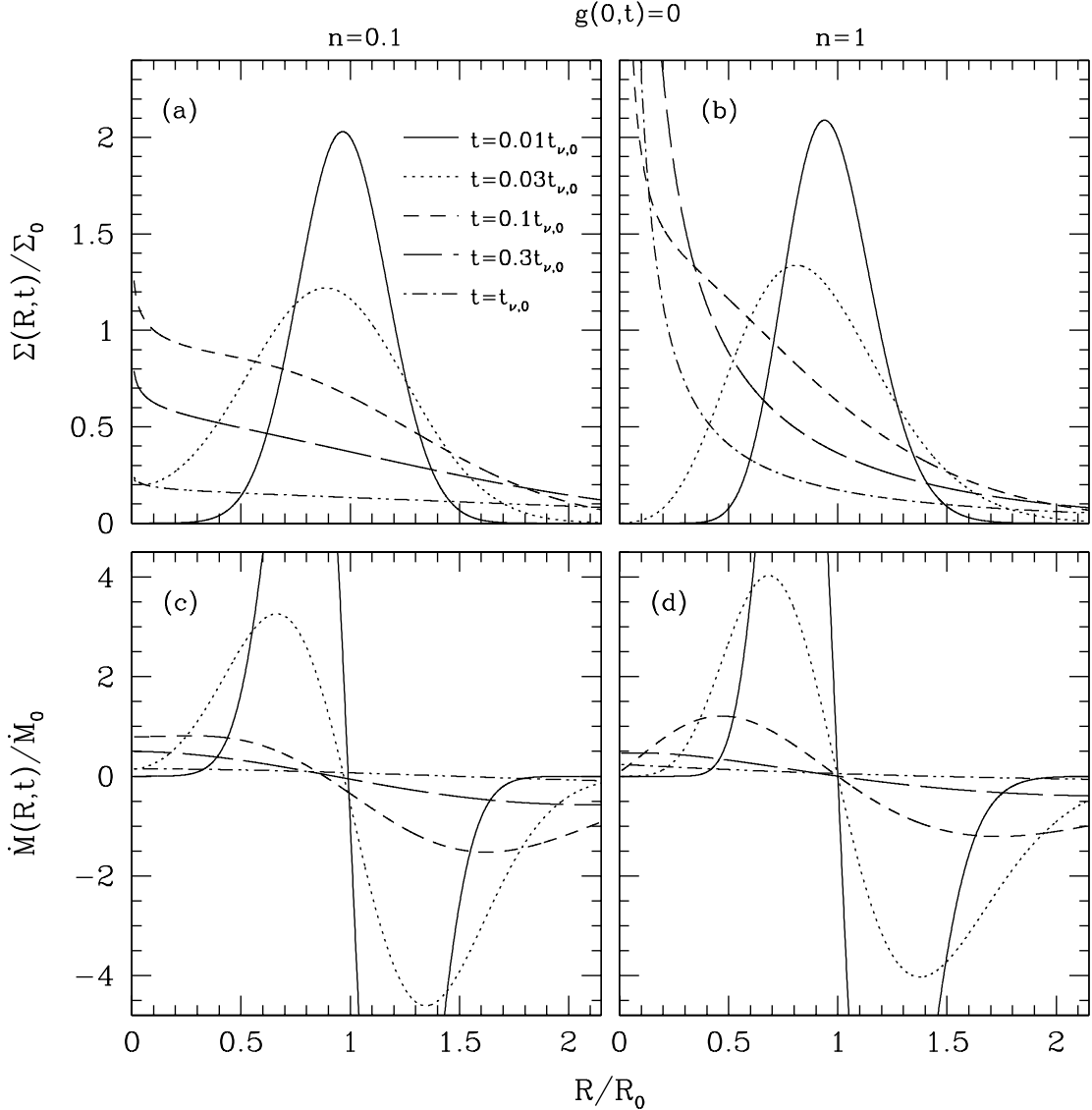, width=15.5cm}
\caption{
The solution $\Sigma(R,t)$, from \citetalias{LP74}, and the corresponding radial mass inflow rate $\dot{M}(R,t)$,
for a zero-torque boundary condition imposed at $R=0$ and a $\delta$-function
initial profile $\Sigma(R,t=0)=\Sigma_{0}R_{0}\delta(R-R_{0})$,
where the quantities $\Sigma_{0}$ and $R_{0}$ are arbitrary.
The viscosity is a radial power law with $\nu\propto R^{n}$.
Panels on the left side (a and c) show solutions for $n=0.1$,
and those on the right (b and d) show solutions for $n=1$.
Values for $t$ are in units of the viscous time at $R_{0}$, $t_{\nu,0}
=(2/3)R_{0}^{2}/\nu(R_{0})$.
We have normalized $\dot{M}$ to the quantity $\dot{M}_{0}\equiv 3\pi \nu(R_{0}) \Sigma_{0}$.
At late times, the solution has the behavior $\Sigma\propto R^{-n}$
and the mass-flow profile $\dot{M}$ becomes flat near the origin.
}
\label{fig:1}
\end{figure}

In Figure \ref{fig:1}, we plot the solution $\Sigma(R,t)$
and the radial mass flow $\dot{M}(R,t)$,
for the $\delta$-function initial condition (equation \ref{eq:delta}),
and for viscosity power-law index values $n=0.1$ and $n=1$.
In both cases, we see the power-law behavior from equation \eqref{eq:asym1}
near the origin as the solution approaches $t\sim t_{\nu,0}$.
The disc spreads as the gas at inner annuli
loses angular momentum to the gas at outer annuli.
The gas initially accumulates near the origin, then
becomes diffuse as mass is lost into the origin.

\subsection{Zero mass flow at $R_{\rm in}=0$}
If the accretion flow has a sufficiently
strong central source of angular momentum,
then the gas will be unable to flow in,
and instead accumulate near the origin.
Such solutions can be used to describe astrophysical discs around
a compact binary \citep{Pringle91},
and perhaps those around
compact objects with strong central magnetic fields \citepalias{LP74}.
For circumbinary thin discs,
\cite{Pringle91} demonstrated that such
a boundary condition characterizes quite well the
effects of an explicit central torque term.

In general, the mass flow has the behavior
\beq
\dot{M}\propto \int_{0}^{\infty}R^{1/2}\frac{\dd}{\dd R}\left[ A(k)J_{\ell}(ky)R^{1/4}+B(k)Y_{\ell}(ky)R^{1/4}\right]~\exp(-3sk^{2}t)~dk.
\eeq
We have seen above
that for solutions with $B(k)=0$ the mass flow is radially constant and finite
near the origin.  On the other hand, because $Y_{\ell}(ky)\propto R^{-1/4}$ near the origin,
the weights $B(k)$ will all contribute no mass flow there;
so for zero mass flow at $R_{\rm in}=0$, we require $A(k)=0$.

We note that because the surface density will have a power-law
$\Sigma \propto R^{-1/2-n}$ at the origin,
for the mass contained in the disc to converge $n$ must be less than $3/2$.
Thus, for physically realistic solutions with zero mass flow
at the origin, $\ell$ cannot be an integer.
It follows that in this case $Y_{\ell}$ in equation \eqref{eq:gensol}
may be replaced by $J_{-\ell}$ without loss of generality. 
Then the Green's function for this case is derived in exactly
the same fashion as in the previous case, the only difference being that
the order of the Hankel transforms has the opposite sign.
We obtain:
\beq
G(R,R^{\prime},t)=(2-n)\frac{R^{-9/4}R^{\prime 5/4}}{\tau(R)}
I_{-\ell}\left[\frac{2\left(R^{\prime}/R\right)^{1-n/2}}{\tau(R)}\right]
\exp\left[-\frac{1+\left(R^{\prime}/R\right)^{2-n}}{\tau(R)}\right].
\eeq

As before, we evaluate the late-time behavior
for the $\delta$-function initial condition (equation \ref{eq:delta}) at small radii:
\beq
\Sigma\left(R, t\ga t_{\nu}(R)\right)\approx
\frac{2-n}{ \Gamma(1-\ell)}\Sigma_{0}
\left(\frac{R}{R_{0}}\right)^{-n-1/2}\left[8\left(1-\frac{n}{2}\right)^{2}\frac{t}{t_{\nu,0}}\right]^{-1+\ell}.
\label{eq:asym2}
\eeq
From the above expression it is clear that the boundary
condition is satisfied: $\dot{M}\propto \dd(\nu\Sigma R^{1/2})\rightarrow 0$
in the limit $R\rightarrow 0$.
Just as we found for the zero-torque boundary condition,
the formal expression for the
power dissipated at the disc center diverges for the zero-flux solution, with
$L(R\le R_{0}, t\ga t_{\nu,0})\propto \int_{0}^{R_{0}}R^{-5/2}~dR$.

The asymptotic behavior at early times and large radii is unaffected
by the order of the function $I_{\ell}$; it is described by equation \ref{eq:Gearly}.
Indeed, the inner boundary condition should have no effect on the disc
at large radii.

\begin{figure}
\centering
\epsfig{file=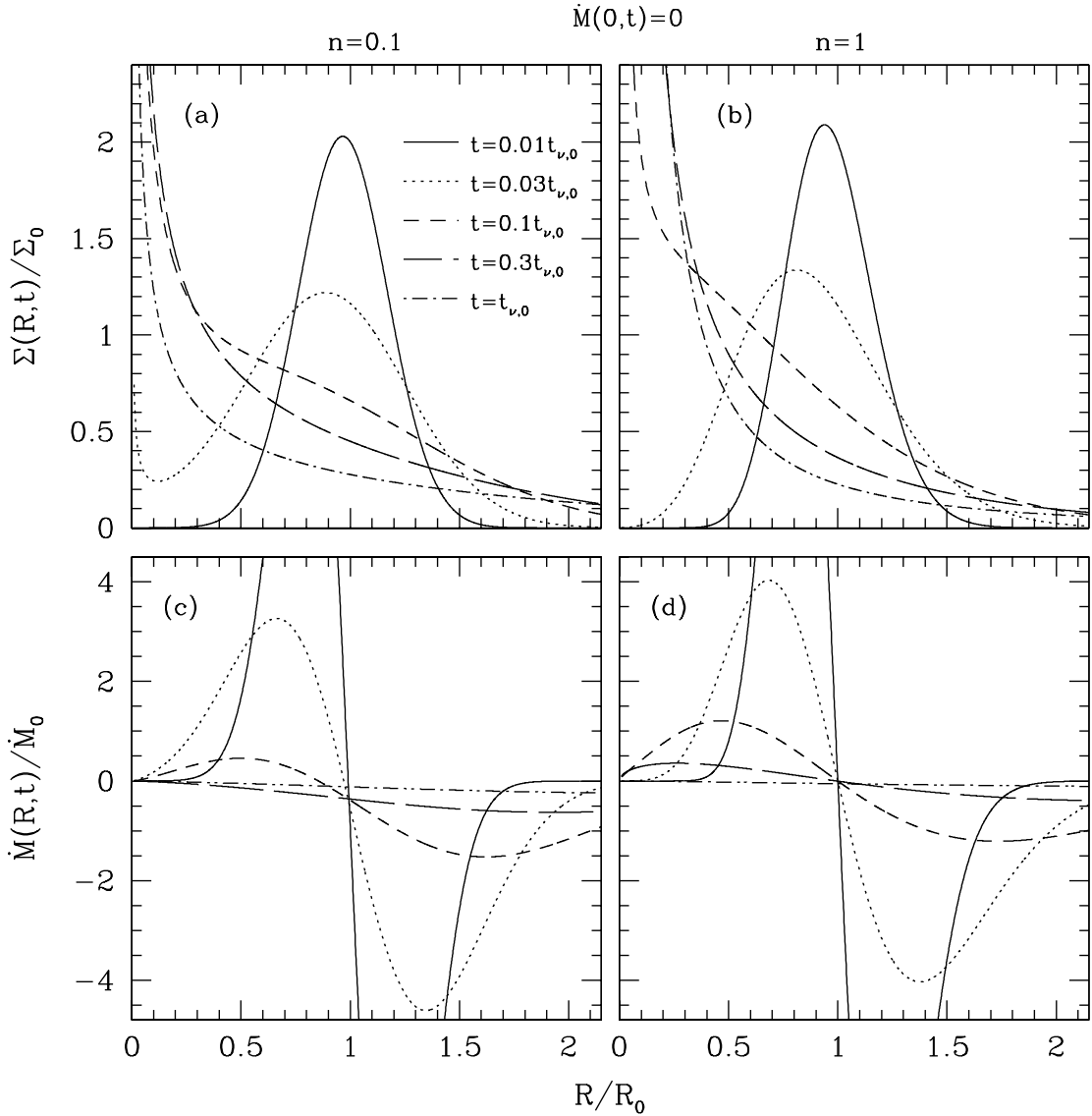, width=15.5cm}
\caption{
Same as Figure \ref{fig:1}, except that the boundary condition is
$\dot{M}=0$ at $R=0$.  Again, the scales $\Sigma_{0}$ and $R_{0}$
are arbitrary.
Whereas in the zero-torque case the total mass in the disc monotonically decreases due to mass loss at the origin
(onto the black hole or star),
the solutions in this figure conserve mass.
At late times, the solution has the behavior $\Sigma\propto R^{-n-1/2}$.
Gas initially piles up near the origin
because of the boundary condition
before gradually spreading outward; note that $\Sigma$ at inner radii decreases
from $t=0.3t_{\nu,0}$ to $t=t_{\nu,0}$.
}
\label{fig:2}
\end{figure}

Figure \ref{fig:2} shows the evolution of the surface density and the radial mass flow
for the boundary condition $\dot{M}(R=0)=0$.
At early times, the behavior is nearly identical to the zero-torque boundary case.
At late times, the zero-flux boundary condition causes the gas to accumulate
instead of being lost to the origin.  The central mass concentration
reaches a maximum, then decreases as the disc begins to spread outward.

\section{Green's-Function Solutions with Boundary Conditions at Finite Radii}
As we have seen above, Green's-function solutions
of thin accretion discs with $R_{\rm in}=0$ have divergent
expressions for the dissipated power, and thus
the innermost surface density profile must be manipulated
to obtain physically realistic disc luminosities.
Analytic treatment of the case with finite boundary radius
was briefly discussed in \citetalias{LP74} and
\cite{Pringle91}, but to the author's knowledge
explicit solutions have never before been published.
We show below that the Green's functions
for finite boundary radii can be derived
with the aid of the appropriate integral transform techniques,
and that they can be represented as ordinary integrals of analytic functions.

\subsection{Zero Torque at $R_{\rm in}>0$}
\label{ssec:new1}
We wish to solve the problem as in \S \ref{subsec:zerotorqueR0}, but with $R_{\rm in}>0$,
i.e.
\beq
g(R_{\rm in})\propto \Sigma\nu R^{1/2}\Big |_{R=R_{\rm in}} \propto \Sigma (R_{\rm in})R_{\rm in}^{n+1/2}=0.
\eeq

We may relate the mode weights $A(k)$ and $B(k)$
by requiring that every mode of the solution satisfy the boundary condition, i.e.:
\beq
A(k)J_{\ell}(ky_{\rm in})+B(k)Y_{\ell}(ky_{\rm in})=0,
\eeq
where $y_{\rm in}\equiv y(R_{\rm in})$.
Substituting $C(k)=A(k)/Y_{\ell}(ky_{\rm in}) = -B(k)/J_{\ell}(ky_{\rm in})$,
we obtain
\beq
\Sigma(R,t)=\int_{0}^{\infty}C(k)R^{-n-1/4}\left[J_{\ell}(ky)Y_{\ell}(ky_{\rm in})-Y_{\ell}(ky)J_{\ell}(ky_{\rm in})\right]\exp(-3sk^{2}t) ~dk.
\eeq

The function $C(k)$ may be evaluated with the use of the Weber integral transform \citep{Titchm23}.
A Weber transform pair satisfies
\begin{align}
\phi_{\ell}(x)&=\int_{0}^{\infty}\Phi_{\ell}(\kappa )\frac{J_{\ell}(\kappa x)Y_{\ell}(\kappa )-Y_{\ell}(\kappa x)J_{\ell}(\kappa )}{J_{\ell}^{2}(\kappa )+Y_{\ell}^{2}(\kappa )}~\kappa ~d\kappa ,\label{eq:Web1}\\
\Phi_{\ell}(\kappa )&=\int_{1}^{\infty}\phi_{\ell}(x)\left[J_{\ell}(\kappa x)Y_{\ell}(\kappa )-Y_{\ell}(\kappa x)J_{\ell}(\kappa )\right]~x~dx.\label{eq:Web2}
\end{align}

Proceeding as before, we construct the pair
\begin{align}
R^{n+1/4}\Sigma(R,t=0)&=\int_{0}^{\infty}\left[C(\kappa )\kappa ^{-1}\right]\frac{J_{\ell}(\kappa x)Y_{\ell}(\kappa )-Y_{\ell}(\kappa x)J_{\ell}(\kappa )}{J_{\ell}^{2}(\kappa )+Y_{\ell}^{2}(\kappa )}~\kappa ~d\kappa , \label{eq:WebC1}\\
C(\kappa )\kappa ^{-1}&=\int_{1}^{\infty}\left[R^{n+1/4}\Sigma(R,t=0)\right]\left[J_{\ell}(\kappa x)Y_{\ell}(\kappa )-Y_{\ell}(\kappa x)J_{\ell}(\kappa )\right]~x~dx.\label{eq:WebC2}
\end{align}

Above, we have substituted $x=y/y_{\rm in}\ge 1$ and $\kappa=ky_{\rm in}$.
Note the lower limit of integration in equation \eqref{eq:WebC2} is nonzero
to account for the finite boundary radius.
Combining equations \eqref{eq:WebC1} and \eqref{eq:WebC2} to eliminate $C(\kappa)$,
and inserting the time-dependence factor
$\exp(-3sk^{2}t)=\exp[-2(1-n/2)^{2}\kappa^{2}t/t_{\nu,\rm in}]$
where $t_{\rm \nu, in}=t_{\nu}(R_{\rm in})$,
we obtain our new Green's function:
\begin{align}
G(R,R^{\prime},t)&=\left(1-\frac{n}{2}\right)R^{-n-1/4}R^{\prime 5/4}R_{\rm in}^{n-2}\nonumber\\
&\qquad\times\int_{0}^{\infty}
\frac{\left[J_{\ell}(\kappa x)Y_{\ell}(\kappa )-Y_{\ell}(\kappa x)J_{\ell}(\kappa )\right]
\left[J_{\ell}(\kappa x^{\prime})Y_{\ell}(\kappa )-Y_{\ell}(\kappa x^{\prime})J_{\ell}(\kappa )\right]}{J_{\ell}^{2}(\kappa )+Y_{\ell}^{2}(\kappa )}\nonumber\\
&\qquad\times
\exp\left[-2\left(1-\frac{n}{2}\right)^{2}\kappa^{2}\frac{t}{t_{\nu,\rm in}}\right]~\kappa ~d\kappa .
\label{eq:G3}
\end{align}
Whereas the integral over $k$ in equation \eqref{eq:G1} has an analytic solution,
to the author's knowledge there is no analytic expression for the integral in equation \eqref{eq:G3}.
Nonetheless, equation \eqref{eq:G3} gives an exact expression for the Green's function.
While it is somewhat more unwieldy than the solutions for $R_{\rm in}=0$,
the additional computational cost of an ordinary integral
is not likely to be a significant practical barrier, 
e.g. one could tabulate the integral in terms of the quantities $x$, $x^{\prime}$ and $t/t_{\rm \nu, in}$.
The boundary condition
has little effect at large radii, so in practice the behavior far from the boundary is
well approximated by the $R_{\rm in}=0$ solutions.

The Green's function in equation \eqref{eq:G1}
does have a closed-form expression for the special case $n=1$ (i.e., $\ell=1/2$).
As noted by \cite{Pringle91}, in this case
the Bessel functions become easier to handle analytically,
with $J_{1/2}(x)=\sqrt{\pi/2}~x^{-1/2}\sin x$ and $Y_{1/2}(x)=-\sqrt{\pi/2}~x^{-1/2}\cos x$.
For this value of $n$ we obtain for our Green's function
\begin{align}
G(R,R^{\prime},t)&=\frac{1}{\pi R_{\rm in}}\left(\frac{R^{\prime}}{R}\right)^{5/4}\left(x~x^{\prime}\right)^{-1/2}
 \int_{0}^{\infty}
\sin\left[\kappa(x-1)\right]\sin\left[\kappa(x^{\prime}-1)\right]
~\exp\left[-\frac{\kappa^{2}}{2}\frac{t}{t_{\nu,\rm in}}\right]~d\kappa\nonumber\\
&=\frac{R^{-3/2}R^{\prime}R_{\rm in}^{-1/2}}{2\sqrt{2\pi}}\sqrt{\frac{t_{\nu,\rm in}}{t}}
\left\{
\exp\left[-\frac{\left(x-x^{\prime}\right)^{2}}{2}\frac{t}{t_{\nu,\rm in}}\right]
-\exp\left[-\frac{\left(x+x^{\prime}-2\right)^{2}}{2}\frac{t}{t_{\nu,\rm in}}\right]
\right\}.
\label{eq:G3n1}
\end{align}
Note that in the case $n=1$, $x$ and $x^{\prime}$ are simply
$\sqrt{R/R_{\rm in}}$ and $\sqrt{R^{\prime}/R_{\rm in}}$, respectively.

For general values of $n$, we can evaluate the behavior at late times $t\ga t_{\nu,0}>t_{\nu,\rm in}$
by noting that in this regime only the modes $\kappa^{2}\ltsim 1$
contribute to the integral in equation \eqref{eq:G3}.
For the central region $R\ltsim R_{0}$ at late times,
we obtain the following analytic expression
for the $\delta$-function initial condition:
\begin{align}
\Sigma\left(R, t\ga t_{\nu}(R)\right)\nonumber
&\approx \frac{2-n}{2^{1+2\ell}\Gamma^{2}(1+\ell)}\Sigma_{0}
\left(\frac{R}{R_{\rm in}}\right)^{-n}\left(\frac{R_{0}}{R_{\rm in}}\right)^{5/2}
\left(1-\sqrt{\frac{R_{\rm in}}{R} }\right)
\left(\sqrt{\frac{R_{0}}{R_{\rm in}} }-1\right)\nonumber\\
&\qquad\qquad\times\int_{0}^{\infty}
\exp\left[-2\left(1-\frac{n}{2}\right)^{2}\kappa^{2}\frac{t}{t_{\nu,\rm in}}\right]~\kappa^{1+2\ell} ~d\kappa\nonumber\\
&=\frac{2-n}{\Gamma(1+\ell)}\Sigma_{0}
\left(\frac{R}{R_{\rm in}}\right)^{-n}\left(\frac{R_{0}}{R_{\rm in}}\right)^{5/2}
\left(1-\sqrt{\frac{R_{\rm in}}{R} }\right)
\left(\sqrt{\frac{R_{0}}{R_{\rm in}} }-1\right)
\left[8\left(1-\frac{n}{2}\right)^{2}\frac{t}{t_{\nu,\rm in}}\right]^{-1-\ell}.
\label{eq:center}
\end{align}
We see that the Green's function explicitly gives the asymptotic behavior
$\Sigma\propto R^{-n}(1-\sqrt{R_{\rm in}/R})$, which has been used extensively
for solutions of accretion discs near zero-torque boundary surfaces
(e.g., \citetalias{LP74}, \citealt{FKR02}).\footnote{The factor arises from
assuming that $\Omega$ is nearly Keplerian at the radius where
the torque $g\propto \dd \Omega/\dd R=0$ \citep{FKR02}.}
This behavior near the boundary and at late times is general for any initial
surface density profile; it is insensitive to the values of $\Sigma_{0}$
and $R_{0}$, and arises for any nonzero $R_{\rm in}$.
This qualitative difference in the inner disc from the $R_{\rm in}=0$ case
also gives a convergent value for the power dissipated in the central disc.
We obtain:
\beq
L\left(R\le R_{t},t\ga t_{\nu}(R)\right)=
\frac{GM\dot{M}_{\rm ss}}{2R_{\rm in}}\left[1-3\frac{R_{\rm in}}{R_{t}}+2\left(\frac{R_{\rm in}}{R_{t}}\right)^{3/2}\right],
\eeq
where $\dot{M}_{\rm ss}$ is the mass flow quantity in equation \eqref{eq:Mdot},
and $R_{t}$ is the radius where $t=t_{\nu}(R)$, inside which the disc has had sufficient
time to approach the asymptotic solution.
In the limit $R_{0}\gg R_{\rm in}$ and $t\gg t_{\nu,0}$, $\dot{M}_{\rm ss}$ may
be interpreted as the mass supply rate into the center of the disc from arbitrarily large radii.
In this limit the above expression agrees precisely with the standard expression for the luminosity of
a steady-state thin accretion disc.  

\begin{figure}
\centering
\epsfig{file=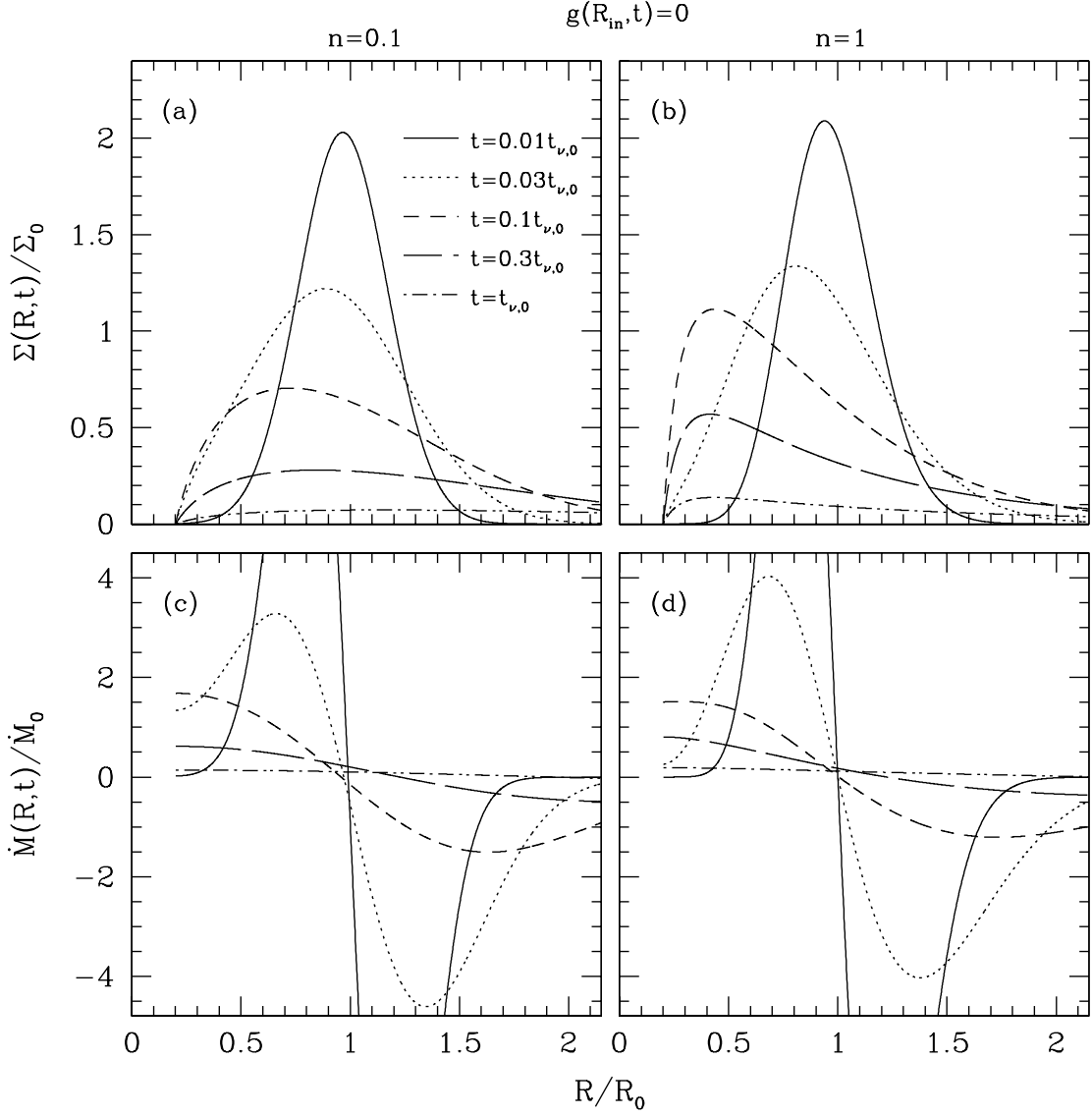, width=15.5cm}
\caption{
Same as as Figure \ref{fig:1}, except that the zero-torque boundary condition is
applied at a finite radius $R_{\rm in}=R_{0}/5$.  As gas
flows near the inner boundary, it exhibits the well-known
behavior $\Sigma\propto R^{-n}(1-\sqrt{R_{\rm in}/R})$
of \citetalias{LP74}.
}
\label{fig:3}
\end{figure}

We show in Figure \ref{fig:3} the exact solutions for the $\delta$-function initial condition,
with the no-torque boundary condition imposed at $R_{\rm in}=R_{0}/5$.
The qualitative evolution is as predicted by \citetalias{LP74}:
at early times, far from the boundary, the disc spreads inward in
very much the same manner as the solutions with $R_{\rm in}=0$,
and so the $R_{\rm in}=0$ Green's function suffices;
at late times, once the gas reaches the vicinity of the boundary it
exhibits the behavior $\Sigma\propto R^{-n}(1-\sqrt{R_{\rm in}/R})$ in that neighborhood.

\subsection{Zero Mass Flux at $R_{\rm in}>0$ }
We now consider the boundary condition of zero mass flow at a particular radius,
\beq
\dot{M}(R_{\rm in})\propto \frac{\dd}{\dd R}\left(\nu\Sigma R^{1/2}\right)\Big{|}_{R=R_{\rm in}}\propto
\frac{\dd}{\dd y}\left\{y^{\ell}\left[A(k)J_{\ell}(ky)+B(k)Y_{\ell}(ky)\right]\right\}\Big{|}_{y=y_{\rm in}}
=0.
\eeq
From the relations $\dd [x^{\ell}J_{\ell}(x)]/\dd x=x^{\ell}J_{\ell-1}(x)$
and $\dd [x^{\ell}Y_{\ell}(x)]/\dd x=x^{\ell}Y_{\ell-1}(x)$,
we obtain the relationship between $A$ and $B$ corresponding to the boundary condition:
\beq
\frac{A(\kappa)}{B(\kappa)}
=
-\frac{Y_{\ell-1}(\kappa)}
{J_{\ell-1}(\kappa)}.
\eeq
The solution is then
\beq
\Sigma(R,t)=\int_{0}^{\infty}C(\kappa)R^{-n-1/4}\left[J_{\ell}(\kappa x)~Y_{\ell-1}(\kappa)-Y_{\ell}(\kappa x)~J_{\ell-1}(\kappa)\right]
\exp\left[-2\left(1-\frac{n}{2}\right)^{2}\kappa^{2}\frac{t}{t_{\nu,\rm in}}\right]~\kappa ~d\kappa
\eeq
\cite{Pringle91} solved the special case $n=1$ analytically, and
noted the mathematical difficulty in deriving a solution for a more general case.
We find that the mode weight $C(\kappa)$ can in fact be solved for with the use of the recently
proved generalized Weber transform \citep{ZT07},
\begin{align}
\phi_{\ell}(x)&=\int_{0}^{\infty}\frac{W_{\ell}(\kappa ,x; a, b)}{Q^{2}_{\ell}(\kappa ; a, b)}
~\Phi_{\ell}(\kappa )~\kappa ~d\kappa ,\\
\Phi_{\ell}(\kappa )&=\int_{1}^{\infty}W_{\ell}(\kappa ,x; a, b)~\phi_{\ell}(x)~x~dx.
\end{align}
The functions $W_{\ell}(\kappa,x; a, b)$ and $Q_{\ell}^{2}(\kappa; a, b)$ are defined as follows:
\begin{align}
W_{\ell}(\kappa ,x; a, b)&\equiv J_{\ell}(\kappa x)\left[a Y_{\ell}(\kappa )+b\kappa  ~Y_{\ell}^{\prime}(\kappa )\right]
-
Y_{\ell}(\kappa x)\left[a J_{\ell}(\kappa )+b\kappa  ~J_{\ell}^{\prime}(\kappa )\right]\nonumber\\
&=    J_{\ell}(\kappa x)\left[\left(a-\ell b\right) Y_{\ell}(\kappa )+b\kappa  ~Y_{\ell-1}(\kappa )\right]
 - Y_{\ell}(\kappa x)\left[\left(a-\ell b\right)  J_{\ell}(\kappa )+b\kappa  ~J_{\ell-1}(\kappa )\right] \qquad\\
Q_{\ell}^{2}(\kappa ; a, b)&\equiv \left[a Y_{\ell}(\kappa )+b\kappa  ~Y_{\ell}^{\prime}(\kappa )\right]^{2}
+\left[a J_{\ell}(\kappa )+b\kappa  ~J_{\ell}^{\prime}(\kappa )\right]^{2}\nonumber\\
&=\left[\left(a-\ell b\right) Y_{\ell}(\kappa )+b\kappa  ~Y_{\ell-1}(\kappa )\right]^{2}
+\left[\left(a-\ell b\right) J_{\ell}(\kappa )+b\kappa  ~J_{\ell-1}(\kappa )\right]^{2}.
\end{align}
Above, $J_{\ell}^{\prime}$ and $Y_{\ell}^{\prime}$ are the ordinary
derivatives of the Bessel functions.
If $a=1$ and $b=0$, the pair is identical to the ordinary Weber transform
(equations \ref{eq:Web1} and \ref{eq:Web2}).

The choice $a=\ell$ and $b=1$ corresponds to the desired boundary condition $\dot{M}(R_{\rm in},t)=0$.
The Green's function is then:
\begin{align}
G(R,R^{\prime},t)&=\left(1-\frac{n}{2}\right)R^{-n-1/4}R^{\prime 5/4}R_{\rm in}^{n-2}\nonumber\\
&\qquad\times\int_{0}^{\infty}
\frac{\left[J_{\ell}(\kappa x)Y_{\ell-1}(\kappa )-Y_{\ell}(\kappa x)J_{\ell-1}(\kappa )\right]
\left[J_{\ell}(\kappa x^{\prime})Y_{\ell-1}(\kappa )-Y_{\ell}(\kappa x^{\prime})J_{\ell-1}(\kappa )\right]}{J_{\ell-1}^{2}(\kappa )+Y_{\ell-1}^{2}(\kappa )}\nonumber\\
&\qquad\times\exp\left[-2\left(1-\frac{n}{2}\right)^{2}\kappa^{2}\frac{t}{t_{\nu,\rm in}}\right]~\kappa ~d\kappa .
\label{eq:G4}
\end{align}

A specific instance of the above Green's function
was derived by \cite{Pringle91} for the case $n=1$.
We can use equation \eqref{eq:G4} to reproduce that previous
solution by noting that $J_{-1/2}(x)=-Y_{1/2}(x)=\sqrt{\pi/2}~x^{-1/2}\cos x$
and $Y_{-1/2}(x)=J_{1/2}(x)=\sqrt{\pi/2}~x^{-1/2}\sin x$.  We obtain:
\begin{align}
G(R,R^{\prime},t)&=\frac{1}{\pi R_{\rm in}}\left(\frac{R^{\prime}}{R}\right)^{5/4}\left(x~x^{\prime}\right)^{-1/2}
 \int_{0}^{\infty}
\cos\left[\kappa(x-1)\right]\cos\left[\kappa(x^{\prime}-1)\right]
~\exp\left[-\frac{\kappa^{2}}{2}\frac{t}{t_{\nu,\rm in}}\right]~d\kappa\nonumber\\
&=\frac{R^{-3/2}R^{\prime}R_{\rm in}^{-1/2}}{2\sqrt{2\pi}}\sqrt{\frac{t_{\nu,\rm in}}{t}}
\left\{
\exp\left[-\frac{\left(x-x^{\prime}\right)^{2}}{2}\frac{t}{t_{\nu,\rm in}}\right]
+\exp\left[-\frac{\left(x+x^{\prime}-2\right)^{2}}{2}\frac{t}{t_{\nu,\rm in}}\right]
\right\}.
\label{eq:G4n1}
\end{align}
The only difference between this Green's function and the one for
$n=1$ and zero torque at $R_{\rm in}$ (equation \ref{eq:G3n1})
is the sign in between the exponential functions.

For general values of $n$, the analytic late-time behavior
of equation \eqref{eq:G4} turns out to be identical to that for the case $R_{\rm in}=0$ (equation \ref{eq:asym2}).
This can be confirmed by observing that for $\ell<1$
and small arguments $\kappa\ll 1$ and $\kappa x\ll 1$,
$W_{\ell}(\kappa, x; \ell, 1)\approx \csc(\ell\pi)J_{-\ell}(\kappa x)J_{\ell-1}(\kappa)$
and $Q_{\ell}^{2}(\kappa; \ell, 1)\approx \csc^{2}(\ell\pi)J_{\ell-1}^{2}(\kappa)$,
and therefore the large fraction in equation \eqref{eq:G4}
is approximately equal to $J_{-\ell}(\kappa x)J_{-\ell}(\kappa x^{\prime})$.
However, because the disc does not extend to the origin for a finite boundary, the integral
for the central disc luminosity converges.  For the $\delta$-function initial surface density profile,
we obtain:
\beq
L\left(R\le R_{t}, t\ga t_{\nu}(R)\right)\sim\frac{GM\dot{M}_{\rm ss}}{R_{\rm in}}
\left(\sqrt{\frac{R_{t}}{R_{\rm in}}}-\frac{R_{\rm in}}{R_{t}}\right)\left[8\left(1-\frac{n}{2}\right)^{2}\frac{t}{t_{\nu,\rm in}}\right]^{\ell},
\eeq
where again $\dot{M}_{\rm ss}$ is the mass supply expression defined in \S\ref{ssec:new1},
and $R_{t}$ is the radius where $t=t_{\nu}(R)$, inside which the disc has had sufficient
time to approach the asymptotic solution.
The above expression for the disc luminosity
is in agreement with the estimate of \cite{IPP99}, who considered
a zero-flux boundary condition in the context of a thin disc around a supermassive black hole binary.

\begin{figure}
\centering
\epsfig{file=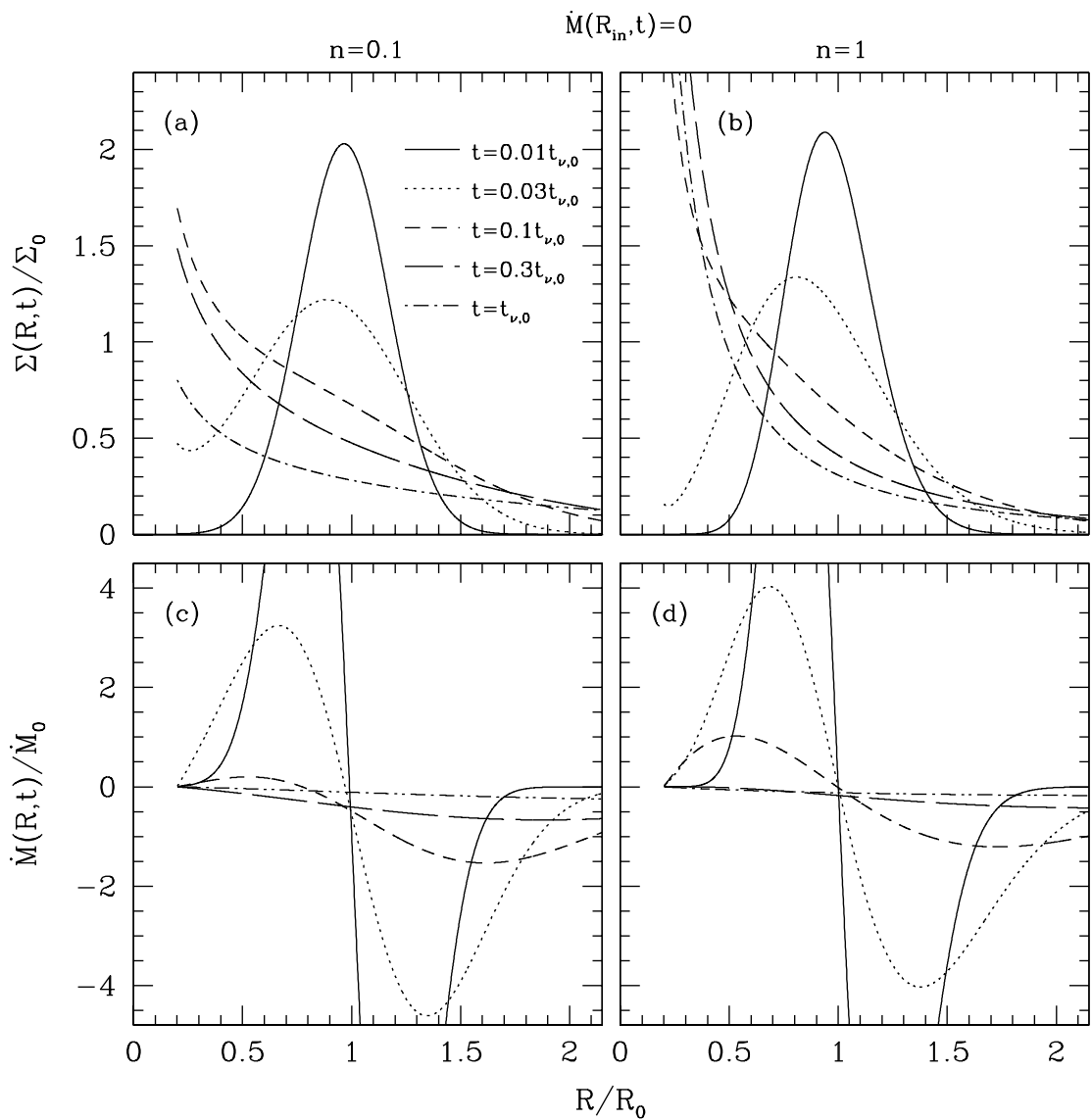, width=15.5cm}
\caption{
Same as as Figure \ref{fig:2}, except that the zero-flux boundary condition is
applied at a finite radius $R_{\rm in}=R_{0}/5$.  Note
that the $n=1$ case was solved analytically by \cite{Pringle91}.
}
\label{fig:4}
\end{figure}

Figure \ref{fig:4} shows the solution for the $\delta$-function initial condition
and the zero-flux boundary condition at $R_{\rm in}=R_{0}/5$.
The panels showing the mass flow clearly exhibit the desired boundary condition.
Note that the case $n=1$ (panels b and d) is the case solved analytically by \cite{Pringle91}.
The $n=1$ case, however, leads to a more rapid evolution
and steeper late-time profiles than solutions with lower values for $n$;
e.g., for the innermost regions of circumbinary discs
around supermassive black holes, the viscosity is believed to
be roughly constant with radius \citep{MP05, TM10}.

\section{Conclusion}
We have presented Green's functions to the equation for viscous
diffusion in a thin Keplerian accretion disc, in the special case of a power-law viscosity profile
$\nu\propto R^{n}$,
for two different types of boundary conditions,
zero viscous torque or zero mass flow, imposed at a finite inner radius $R_{\rm in}>0$.
They are extensions of the elegant analytic solutions derived by \cite{Lust52} and
\citetalias{LP74}
for the same boundary conditions applied at $R_{\rm in}=0$.
While the problem of the finite-radius boundary had been
mentioned previously in the literature, to the author's knowledge
these solutions have not been explicitly pursued,
and are presented here for the first time.
The new solutions can be used to model the time-dependent
behavior of the innermost regions of accretion discs,
where the finite physical size of the central objects
can significantly affect the observable characteristics
of the disc.
Whereas the power viscously dissipated in the $R_{\rm in}=0$
solutions diverge, and require manipulation of the profile
at the disc center to calculate physically plausible disc luminosities,
the power for the new solutions converge to expressions
that are consistent with disc luminosities inferred by other (non-Green's function)
methods.
The solutions presented here complement the numerous approximate solutions and numerical
treatments in the literature.\footnote{For example,
\cite{Cannizzo+90} studied the accretion of a tidally disrupted star onto a black hole
via numerical solutions and analytic self-similar solutions.
The problem of a thin disc with $\dot{M}=0$ at a finite radius
was discussed for the non-linear case $\nu\propto \Sigma^{m}\nu^{n}$
by \cite{Pringle91} and \cite{IPP99}, with both papers
providing numerical solutions as well as analytic approximations.}

The integral transforms used to derive the solutions
are applicable to a wide class of boundary conditions,
and may be applicable to astrophysical thin-disc systems and configurations
not considered here.  Because the generalized Weber transform
by its nature is applicable to many second-order differential equations
with intrinsic cylindrical symmetry, they may also prove to be useful
in solving other mathematical equations in astrophysics and other fields.

\section*{Acknowledgements}
It is a pleasure to thank Kristen Menou and Zolt\'an Haiman for helpful
conversations and comments on the manuscript;
and Jim Pringle and Jeremy Goodman for consultation regarding the literature.
The author is also grateful to the Kavli Institute for Theoretical Physics,
where a part of this work took place, for their hospitality.
Support for this work was provided by NASA ATFP grant NNXO8AH35G (to KM and ZH),
and also by the Pol\'anyi Program of the Hungarian National Office of Technology (to ZH).
\bibliographystyle{apj}
\bibliography{paper}

\label{lastpage}
\end{document}